\documentclass[a4paper]{article}

\usepackage[english]{babel}
\usepackage[utf8x]{inputenc}
\usepackage[T1]{fontenc}

\usepackage[a4paper,top=3cm,bottom=2cm,left=3cm,right=3cm,marginparwidth=1.75cm]{geometry}

\usepackage{amsmath,amsxtra}
\usepackage{graphicx}
\usepackage[colorinlistoftodos]{todonotes}
\usepackage[colorlinks=true, allcolors=blue]{hyperref}

\begin{document}

\title{Classification of cryptocurrency coins and tokens by the dynamics of their market capitalisations}

 \author{Ke Wu\footnote{Corresponding author. Address: Chair of Entrepreneurial Risks at ETH Zurich, Scheuchzerstrasse 7, 8092 Zurich, Switzerland, Email: kwu@student.ethz.ch} \thanks{\rm  ETH Z\"{u}rich, Dept. of Management, Technology and Economics, Z\"{u}rich, Switzerland} , 
 Spencer Wheatley\footnotemark[1]   ~ and Didier Sornette\footnotemark[1] \thanks{\rm also at the Swiss Finance Institute, c/o University of Geneva, Switzerland}}

\maketitle

\begin{abstract}
We empirically verify that the market capitalisations of coins and tokens in the cryptocurrency universe follow power-law distributions with significantly different values, with the tail exponent falling between 0.5 and 0.7 for coins, and between 1.0 and 1.3 for tokens. We provide a rationale for this, based on a simple proportional growth with birth \& death model previously employed to describe the size distribution of firms, cities, webpages, etc.  We empirically validate the model and its main predictions, in terms of proportional growth (Gibrat's law) of the coins and tokens. Estimating the main parameters of the model, the theoretical predictions for the power-law exponents of coin and token distributions are in remarkable agreement with the empirical estimations, given the simplicity of the model. Our results clearly characterize coins as being ``entrenched incumbents'' and tokens as an ``explosive immature ecosystem'', largely due to massive and exuberant Initial Coin Offering activity in the token space. The theory predicts that the exponent for tokens should converge to $1$ in the future, reflecting a more reasonable rate of new entrants associated with genuine technological innovations.
\end{abstract}

\vskip 5cm

\pagebreak

\section{Introduction}

In 2008, under the pseudonym Satoshi Nakamoto, the decentralized cryptocurrency, Bitcoin \cite{nakamoto2008bitcoin}, and its innovative and disruptive blockchain technology\footnote{Powered by a public decentralized ledger that records and validates all transactions chronologically, called the blockchain. These transactions are secured and verified by encryption techniques, and shared between network participants in the absence of a central authority.} was introduced. From its techno-libertarian beginnings, Bitcoin, and a host of other cryptocurrencies have turbulently erupted into the mainstream.
In an overall story of tremendous growth, by Feb 2018 around $1500$ cryptocurrencies exist with their total market capitalization 
hitting an all-time high of \$830 billion on Jan 7, 2018, and then crashing to \$280 billion in the following month -- a sensational drop, but only partially undoing gains made in Q4 2017. Growth potential and market action have therefore attracted huge attention among retail and institutional investors, who are rushing into the new ``crypto-world''\footnote{According to Fintech research house Autonomous NEXT, the number of crypto hedge funds more than doubled in the four months to Feb 15, 2018 \cite{keidan2018number}.}, whose hype is based on the key promise that cryptocurrency technology can deliver decentralized systems that avoid trust and reliance upon centralized authorities, and keep power in the hands of the users. A range of disruptive use cases, some more speculative than others, are foreseen\footnote{ Such as being a global decentralized currency, avoiding central banks, and ``banking the un-banked''; a secure digital asset, within the class of safe haven assets, such as gold or perceived stable currencies; and even a fully decentralized internet, whose protocol hosts a wide range of distributed applications.}.
At the same time, well-known figures from central banks, governments, financial institutions, and other status quo agents, have censured the cryptocurrency space -- calling it a ``scam'' with zero fundamental value. Unsurprisingly, regulators are watching the space, and their early statements about potential regulation send shock waves through the market. 

Regarding academic studies of cryptocurrency, aside from some comprehensive surveys \cite{caginalp2018opinion,Fantazzini2016}, studies have mostly focused on Bitcoin.
This includes: economics \cite{ali2014economics, vigna2015theage, chiu2017economics, ciaian2016economics}; network properties \cite{teo2015emergence,kondor2014rich,ober2013structure, tasca2016evolution}; social signals \cite{kristoufek2013bitcoin,garcia2014digital,Garcia2015,Glaser2014,Kim2017}  and price dynamics \cite{sapuric2014bitcoin,Pichl2017,Urquhart2018,Bouri2017,Bariviera2017,Osterrieder2017,chu2015statistical,kristoufek2015main,dyhrberg2016bitcoin}. Focusing on overall market dynamics and growth mechanisms, some models have been proposed \cite{bornholdt2014bitcoins,hileman2017global}, but failed to reliably explain the market dynamics. For instance, in \cite{elbahrawy2017evolutionary}, their ecological model predicted a gradual drop of Bitcoin to 50 percent of the total market capitalization in a decade from now, but that same drop then happened within months of the paper being published. 

For guidance, we look to fundamental work on the nature of growth of firms and other entities. In particular, Zipf's Law has been 
identified as an ubiquitous empirical regularity for
firm sizes~\cite{axtell2001zipf}, 
city sizes~\cite{gabaix1999zipf}, 
connections between Web pages~\cite{kong2008experience},
connections between open source software packages~\cite{maillart2008empirical}, etc. -- manifesting as a power law distribution of sizes with a unit parameter, such that $Pr\lbrace Size >x \rbrace \propto x^{-1}$ for sufficiently large size level, x. Since Simon's pioneering work~\cite{simon1955class}, the primary generating mechanism of Zipf's law is understood to be proportional growth (``Gibrat's law''), also popularized as ``preferential attachment'' when recast in the context of networks~\cite{barabasi1999emergence}. 
Malevergne et al.~\cite{malevergne2013zipf} extended the proportional growth framework to feature realistic birth and death, which again yields Zipf's law, but not necessarily with unit parameter -- depending on a balance between the growth of new firms versus old ones. We employ this framework to study the growth process of cryptocurrencies, according to their market capitalization, from April 2013 to Feb 2018 \footnote{This study uses the daily data of 2499 cryptocurrencies from April 28, 2013 until February 7, 2018; 1497 of which are still alive on Feb 7. The data is taken from Coin Market Cap~\cite{coinmarketcap}, including daily closing price, market capitalization (the product of the price and the circulating supply), and the type of the cryptocurrency (coin or token) \cite{dryadData}.}. 
We make an essential distinction that some cryptos are ``coins'' -- which operate on their own independent network -- and others ``tokens'' -- which operate on top of a coin network as a platform. Notably, the coin market capitalization distribution is heavier tailed than Zipf's Law, and that of the token market somewhat lighter. The framework of Malevergne et al.~\cite{malevergne2013zipf} allows this to be explained, and identifies that the coins and tokens currently exist in distinct market regimes. This requires confirming Gibrat's law, estimating the birth and death parameters, and comparing the predicted exponent of the market capitalization distribution with its empirical counterpart, both for coins and tokens. Despite the clear limitations of the model, and a highly non-stationary market, we argue that this provides a reliable and meaningful result which may be refined with extended methods.

\section{Evolution of Crypto-Currencies and Token Market Capitalization}%

\begin{figure}[!htb]
	\centering
	\includegraphics[width=0.7\linewidth]{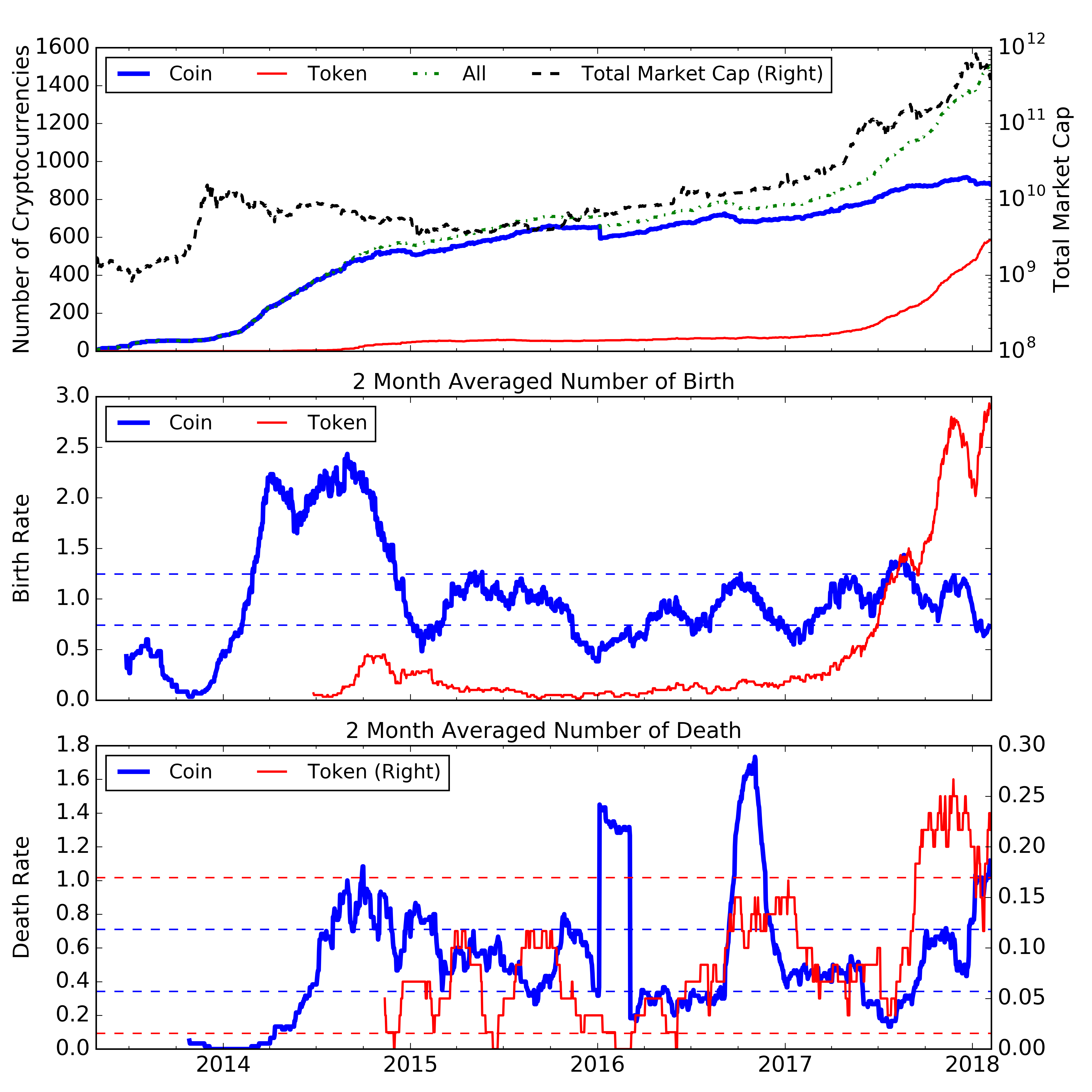}
	\caption{The upper panel depicts the evolution of the number of all cryptocurrencies (green dash dotted line), including both coins (blue thick line), and tokens (red thin line). The total market capitalization of all cryptocurrencies is plotted with a black dashed line against the right y-axis (log scale). The middle panel plots the birth rate for coins (blue thick line), and tokens (thin line) respectively. The lower panel is the corresponding death rate. Birth and death rates are the averaged number of births (deaths) in 2 month moving window. The horizontal blue (resp. red) dashed lines in the middle and lower panels are the 95\% confidence intervals, assuming a Poisson process for coins (resp. tokens), whose mean is estimated over the whole period. The number of births is calculated as the number of new coins/tokens that appeared on CoinMarketCap\cite{coinmarketcap} each day, and the number of deaths is calculated as the number coins/tokens removed from CoinMarketCap (i.e. marked as ``inactive'' by CoinMarketCap)\cite{coinmarketcap}}
	\label{fig:birth_death}
\end{figure}

After going through about a two year bear market, the cryptocurrency market started to grow again at the beginning of 2016 (Figure~\ref{fig:birth_death}). The total market capitalization of all cryptocurrencies achieved a 250\% return in 2016, and 3170\% return in 2017. Although the first token, ``Maid Safe Coin'', appeared in April 2014, 
not until 2017 did the number of tokens explode, from less than 50 to more than 400 by the end of the year. On the other hand, the birth of coins has been relatively stable and the market therefore more mature. The evolution of the number of token deaths is more noisy, in some cases due to external events such as the bankruptcy of a large exchange in 2016~\cite{higgins2016cryptsy}.

\begin{figure}[!htb]
	\centering
	\includegraphics[width=0.9\linewidth]{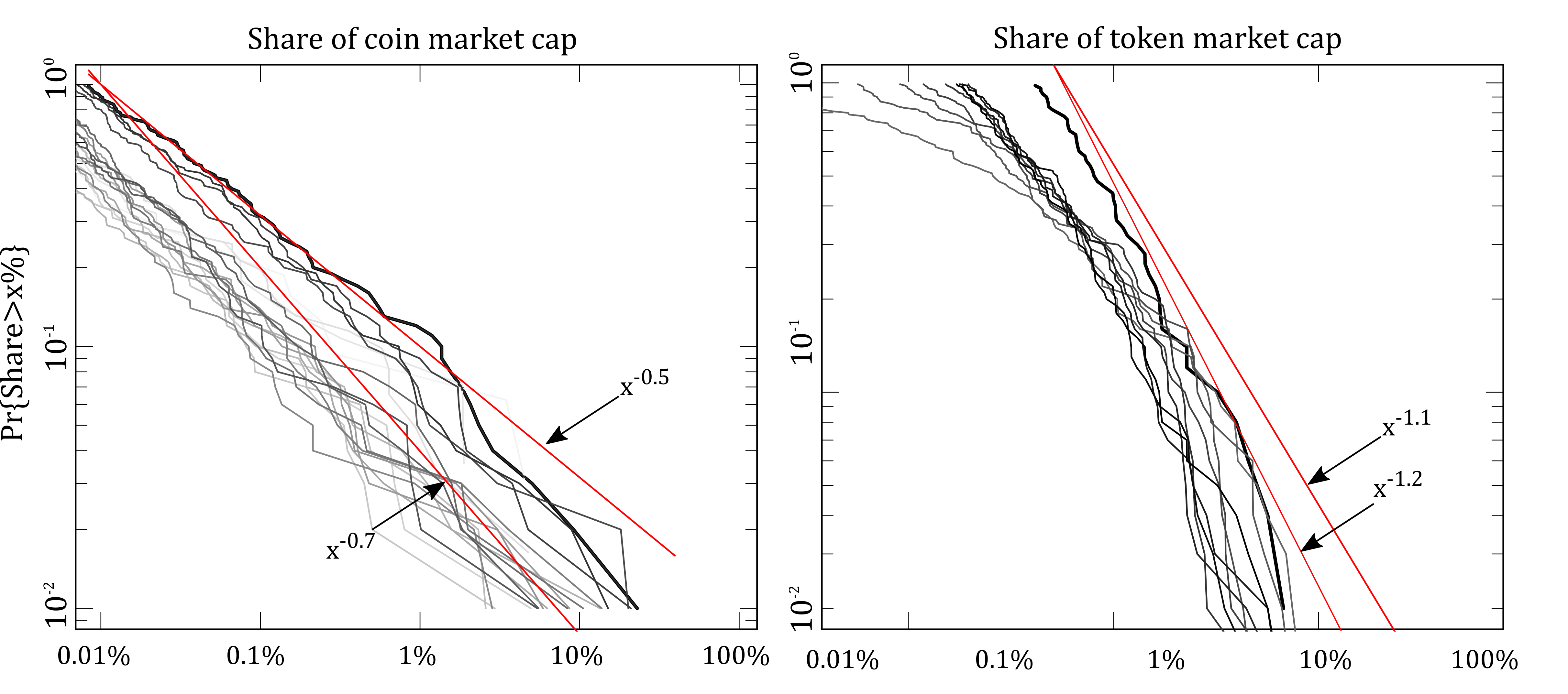}
	\caption{Left: Empirical complementary cumulative distribution function (CCDF) of top 100 coin market capitalizations at snapshots from 2014 (grey) to February 2018 (black) with the range of fitted Pareto tails indicated by the two red lines. Right: The same for tokens, but for the top 50 tokens, starting in early 2017.}
	\label{fig:coin_Token_df}
\end{figure}

Although we are interested in market capitalization, to briefly isolate the relative size of different cryptocurrencies, we examine the distribution of market shares (the fraction of each coin or token to the total market capitalization of all coins or tokens, respectively) in Figure \ref{fig:coin_Token_df}.  For coins, the distribution is well described by a Pareto (``power law'') distribution, 
\begin{equation}
	\label{eqt:powerlaw}
	Pr\lbrace X>x \rbrace = (x/u)^{-\mu} , ~x>u>0.
\end{equation}
with the tail exponent, $\mu>0$, fluctuating between 0.5 and 0.7 over time\footnote{The tail exponent is estimated by Maximum Likelihood}, and not being significantly worse than the more flexible 2-parameter Lognormal distribution for the top 275, out of more than 500 coins, in the most recent snapshot of February 2018 \footnote{Lognormal not superior at p=0.05 level, using uniformly most powerful unbiased test \cite{malevergne2011testing}.}. The market share distribution of tokens was closer to Lognormal instead of power-law at the earlier stage of 2017, but has been recently evolving towards a power-law in the past months, with parameter around 1.1, and not being significantly worse than the Lognormal for the top 50, out of more than 400 tokens, at a 0.05 test level. This confirms that the tail of coin and token market capitalization distributions are now well described by power laws with different exponents. Recall that the Pareto distribution with $\mu=1$ is a border case called Zipf's law \cite{saichev2010} where all moments of order larger than or equal to $1$ are infinite. In the next section, we will consider a model to explain why coin and token distributions fall on different sides of this border case $\mu=1$.

\begin{figure}[!htb]
	\centering
	\includegraphics[width=0.9\linewidth]{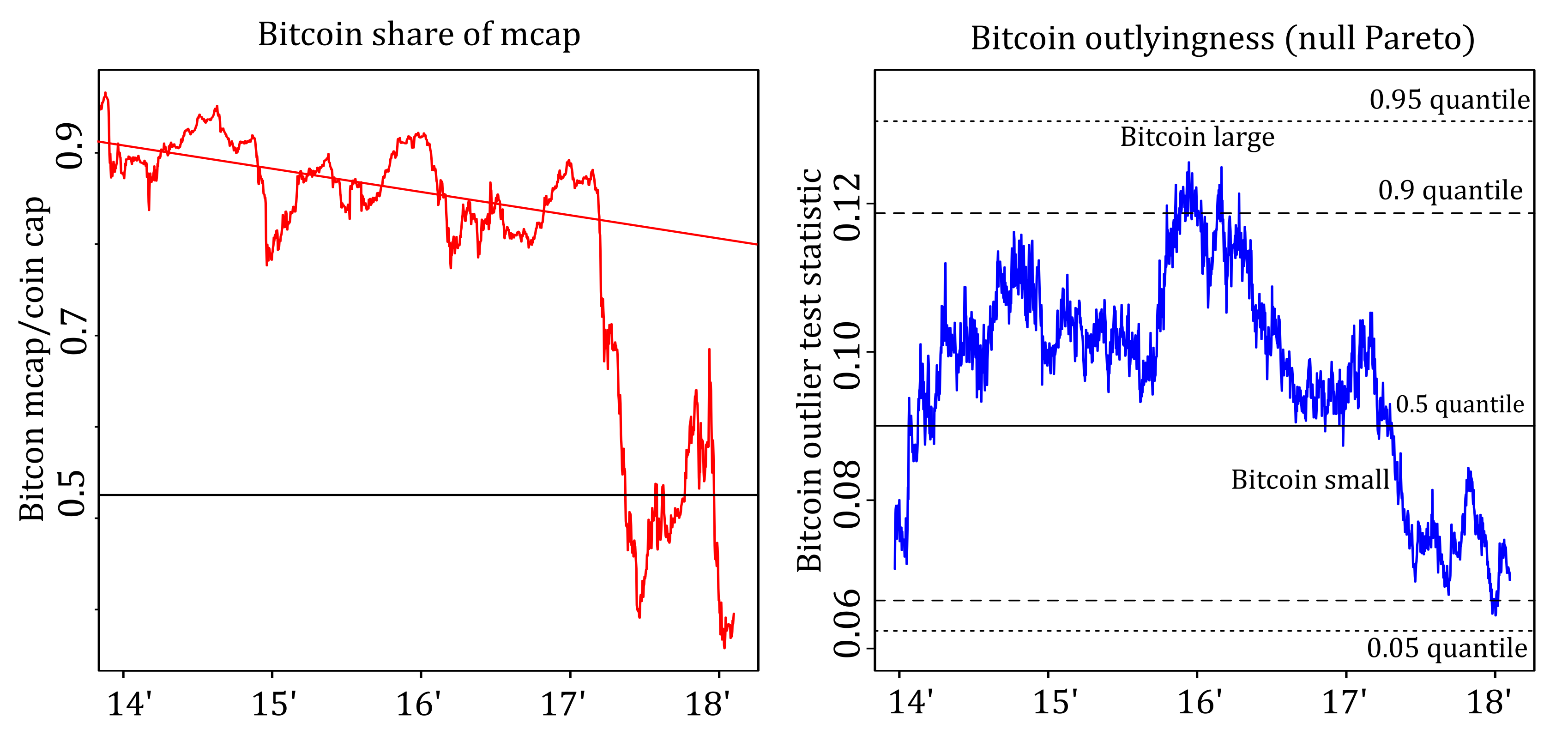}
	\caption{ Left: Bitcoin dominance (share of total coin market capitalization) over time. Right: Test statistic and null (under Pareto) quantiles for testing Bitcoin as an outlier. In particular, the test statistic is $E_1/(E_2+\dots+E_{100})$ where $E_i=log(X_i /X_{100}), i=1,\dots,100$ is a transformation of the top 100 market capitalizations $X_1>\dots>X_{100}$ that transforms $X$ with a Pareto distribution to $E$ with an exponential distribution \cite{wheatley2015multiple}. With this transformation, the test statistic is independent of the parameter. }
	\label{fig:btc_dom}
\end{figure}

Before this, we address the ``Bitcoin maximalism'' belief that Bitcoin would be the one and only winner, and all alternative coins (``altcoins'') are destined to fail.  This degenerate scenario would preclude our growth framework. However, time has largely settled the debate on this, which we briefly address in Figure~\ref{fig:btc_dom}, where Bitcoin dominance (its coin market share) has dropped from above 80 percent to at times well below 50 percent. The follow-up question is then, if Bitcoin is or has been an outlier. Bitcoin dominance alone may be misleading since the market capitalization distribution has changed over time. However, the market size of Bitcoin can be compared to the other top 100 market capitalization coins via a transformation \cite{wheatley2015multiple}. As shown in Figure~\ref{fig:btc_dom}, in 2016, Bitcoin was at its relative largest, at times exceeding the 0.9 quantile of the null distribution. Since then, it has descended to its relative smallest size, near the 0.1 null quantile. This indicates a change of fortune for Bitcoin, and effectively rejects it as an outlier\footnote{Note that here Bitcoin forks are treated as separate independent cryptocurrencies, however including them all together within the Bitcoin value provides similar results}.

\section{Proportional Growth with Stochastic Birth and Death}

\subsection{Definition of the Model and Main Properties}

Proportional growth is a general and ubiquitous mechanism, as discussed in the introduction, and is quite natural for cryptocurrencies given the pervasiveness of proportional growth in complex networks. Within the cryptocurrency community, ``network effects'' have often been attested as a reason for the sustained dominance of Bitcoin. Further, allowing for birth and death, we employ the framework of Malevergne et al. \cite{malevergne2013zipf}, which is based on the following assumptions:

\begin{enumerate}
	\item {\bf Gibrat's rule of proportional growth holds}. This implies that, in the continuous time limit, the market capitalization $MC_i(t)$ of the $i_{th}$ cryptocurrency at time $t$, conditional on its initial market capitalization, is the solution to the stochastic differential equation (i.e. geometric Brownian motion)
	\begin{equation}
		\label{eqt:gbm}
		dMC_i(t)=MC_i(t)\left[rdt+\sigma dW_i(t)\right] ,
	\end{equation}
	where $r$ is the drift and $\sigma$ is the standard deviation, and $W(t)$ is a standard Wiener process. Parameters $r$ and $\sigma$ are assumed to be the same for all cryptocurrencies, but the Wiener process $W_i(t)$ is specific to each.
	\item {\bf Independent random birth time and size}.  The birth flow of each crypto, at time $t_i, i\in N$,  follows a Poisson process with exponentially growing intensity $v(t)=v_0e^{d\cdot t}$, and initial size  $s_0^i=s_{0,i}\cdot e^{c_0t_i}$, where $\{s_{0,i}\}_{i\in N}$ are independent draws from a common random variable.\footnote{Exponential growth is a standard feature of economic systems and financial markets. However, extension of the birth time process to a vast class of non-Poisson processes does not alter key results.} 
	\item {\bf Cryptocurrencies exit (die) at random with a constant hazard rate, $h\geq0$, independent of size}.
\end{enumerate}

Under these assumptions and mild conditions, asymptotically, the process generates a power-law distribution with tail index $\mu_{TH}$
\begin{equation}
	\label{eqt:mu_th}
	\mu_{TH} := \frac{1}{2} \left[ (1-2 \frac{r-c_0}{\sigma^2}) + \sqrt{(1-2 \frac{r-c_0}{\sigma^2})^2 + 8 \frac{d+h}{\sigma^2}} \right] .
\end{equation}

It is important to stress that this is a very simple model with a number of limitations, listed below, and therefore the objective is only to capture the rough fundamental dynamics of the crypto ecology.

\begin{itemize}
	\item It does not capture the strong non-stationarities (e.g., bubbles and crashes) of the market;
	\item It treats cryptocurrencies as independent despite the overall market being highly correlated, including some pairs being more correlated than others;
	\item It neglects difference in ``fitness'' (i.e., quality) of different cryptocurrencies, which has been shown to be important in complex networks \cite{bianconi2001competition}, and is clearly present as newer technology is introduced in newer coins;
	\item It neglects ``forking'', which is similar to spin-off/divestiture of a company;
	\item The process only applies above a sufficiently high threshold, acknowledging that an entire complex ecosystem cannot be described by such a simple model.
\end{itemize}

\subsection{Direct Empirical Quantitative Confirmation of Gibrat's Law of Proportional Growth}

\begin{figure}[!htb]
	\centering
	\includegraphics[width=0.9\linewidth]{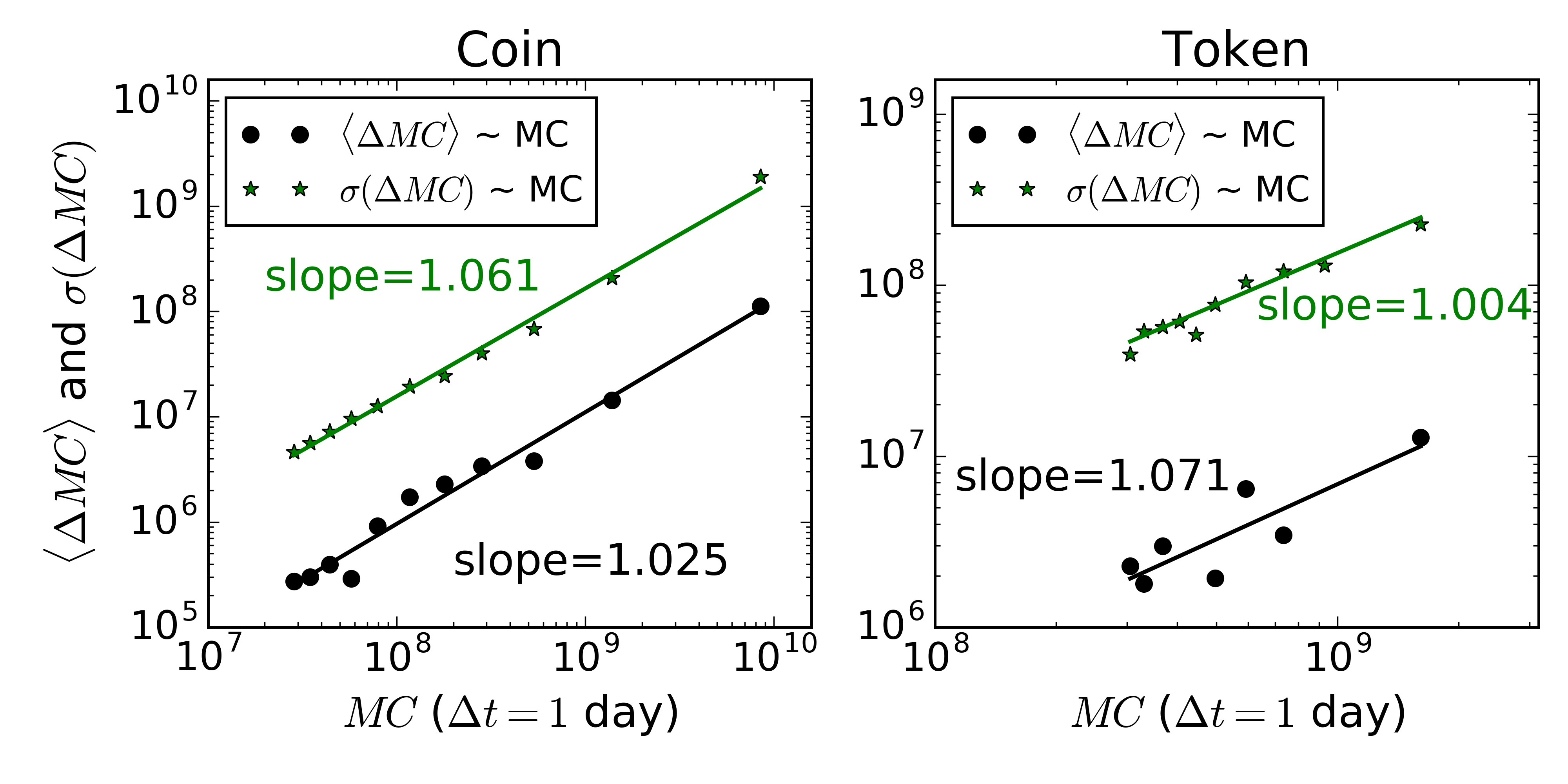}
	\caption{Test of Gibrat's law of proportional growth for market capitalization of large coins (left panel) and tokens (right panel) within a one-year window, up to Feb 7, 2018. The black circles are the mean of the increments (i.e., $\langle \Delta MC \rangle$) versus its current market capitalization $MC$. The green stars are the standard deviation of the increments (i.e., $\sigma(\Delta MC)$) versus its current market capitalization. Every 2000 increments (i.e. $\Delta MC_t$) for coins (resp. 250 for tokens) are grouped into a subset for calculating these means and standard deviations. Only positive points are shown. In both panels, the lines show the least squares fit to the data points.}
	\label{fig:gibrat_rule_linear}
\end{figure}

Gibrat's law of proportional growth embodied in Equation~\ref{eqt:gbm} implies that, for sufficiently small time intervals $\Delta t$, the mean change in market capitalization $\langle \Delta MC \rangle$ and the standard deviation of the change, $\sigma(\Delta MC)$, are both proportional to $MC$ for large coins and tokens. Figure~\ref{fig:gibrat_rule_linear} shows the mean and standard deviation of $\Delta MC$ as a function of $MC$, setting $\Delta t$ as one day, within a one year time window, confirming proportional growth.

Moreover, Equation~\ref{eqt:gbm} implies that, over a small time interval $\Delta t$, the average growth rate $\langle \frac{\Delta MC}{MC} \rangle$ and its standard deviation should be given by,
\begin{equation}
	\langle \frac{\Delta MC}{MC} \rangle = r \times \Delta t ,~~\sigma (\frac{\Delta MC}{MC}) = \sigma \times \sqrt{\Delta t} ,
\end{equation}
where the later square-root dependence reflects the property of the Wiener (random walk) process.

\begin{figure}[!htb]
	\centering
	\includegraphics[width=0.9\linewidth]{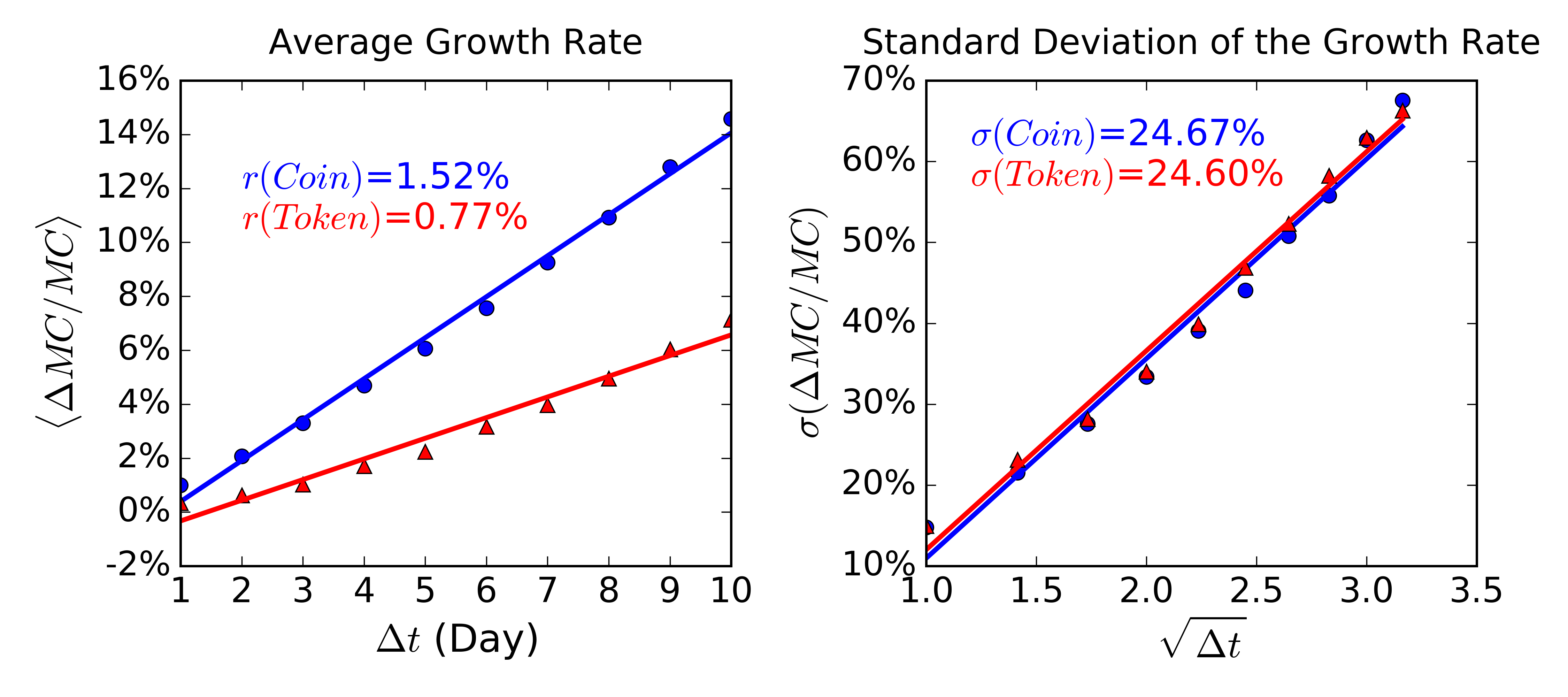}
	\caption{Left: The relationship between the average growth rate $\langle \frac{\Delta MC}{MC} \rangle$ versus the time interval $\Delta t$, for coins (blue circle) and tokens (red triangles) respectively. Right: The standard deviation of the growth rate $\sigma (\frac{\Delta MC}{MC})$ versus $\sqrt{\Delta t}$, for coins and tokens respectively. Data values were taken in the one year window ending Feb 7, 2018.}
	\label{fig:mc_dmc_dt_growthrate}
\end{figure}

This is verified via Figure~\ref{fig:mc_dmc_dt_growthrate}, where we estimate the drift $r$ (resp. the standard deviation $\sigma$) as the slope of the linear regression of the average growth rate (resp. standard deviation of the growth rate) as a function of $\Delta t$ (resp. $\sqrt{\Delta t}$), with $\Delta t\leq 10$ days. We can see that the growth rate of coins is roughly two times that of tokens, while their volatilities are similar -- the relatively large growth in the token market capitalization is therefore a result of the high birth rate, not exceptional growth of individual tokens.

\subsection{Estimation of the Birth and Death Parameters }

\begin{figure}[!htb]
	\centering
	\includegraphics[width=0.9\linewidth]{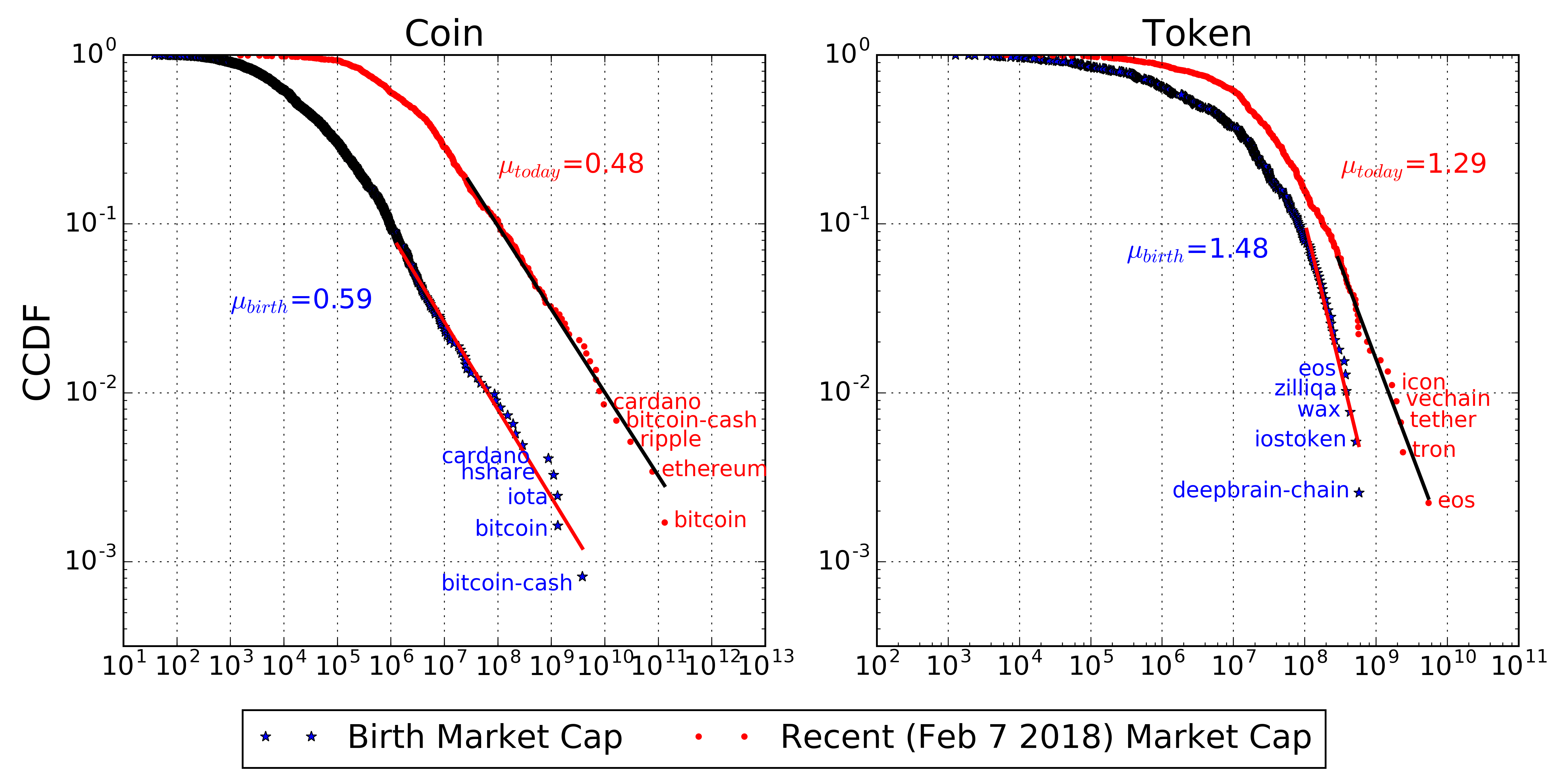}
	\caption{Comparison between the distributions of birth market capitalization and the recent market capitalization, for coins (left panel) and tokens (right panel) respectively. The black stars are the market capitalization at birth (taking the market capitalization 1 week after the birth as a proxy where supply and price are both known). The red dots are the market capitalization on Feb 7, 2018. The tail exponents are estimated by the largest 100 coins and 30 tokens respectively. The largest five coins/tokens are labeled in the upper panel.}
	\label{fig:ccdf_initial_coin_token}
\end{figure}

For both coins and tokens, the distribution of birth market capitalization\footnote{Note that the birth market capitalization we get from CoinMarketCap may have some delays, because: 1) sometimes cryptocurrencies are listed to CoinMarketCap after they have been traded for a while; 2) for some cryptocurrencies, the information of the circulating supply is not available to calculate the market capitalization at the birth time. Therefore, the distribution of market capitalization at birth (``initial market capitalization'') we show in Figure~\ref{fig:ccdf_initial_coin_token} is later than the actual birth time.} (see  Figure~\ref{fig:ccdf_initial_coin_token}) has a substantially thinner tail (0.59 for coins and 1.48 for tokens) than the current distribution (0.48 for coins and 1.29 for tokens), whose exponents are estimated for the largest 100 coins and 30 tokens respectively, based on Maximum Likelihood. This confirms that the observed market capitalization distribution is not simply a consequence of the distribution of initial market capitalizations. Rather, the distributions becomes heavier-tailed due to proportional growth.

\begin{figure}[!htb]
	\centering
	\includegraphics[width=0.9\linewidth]{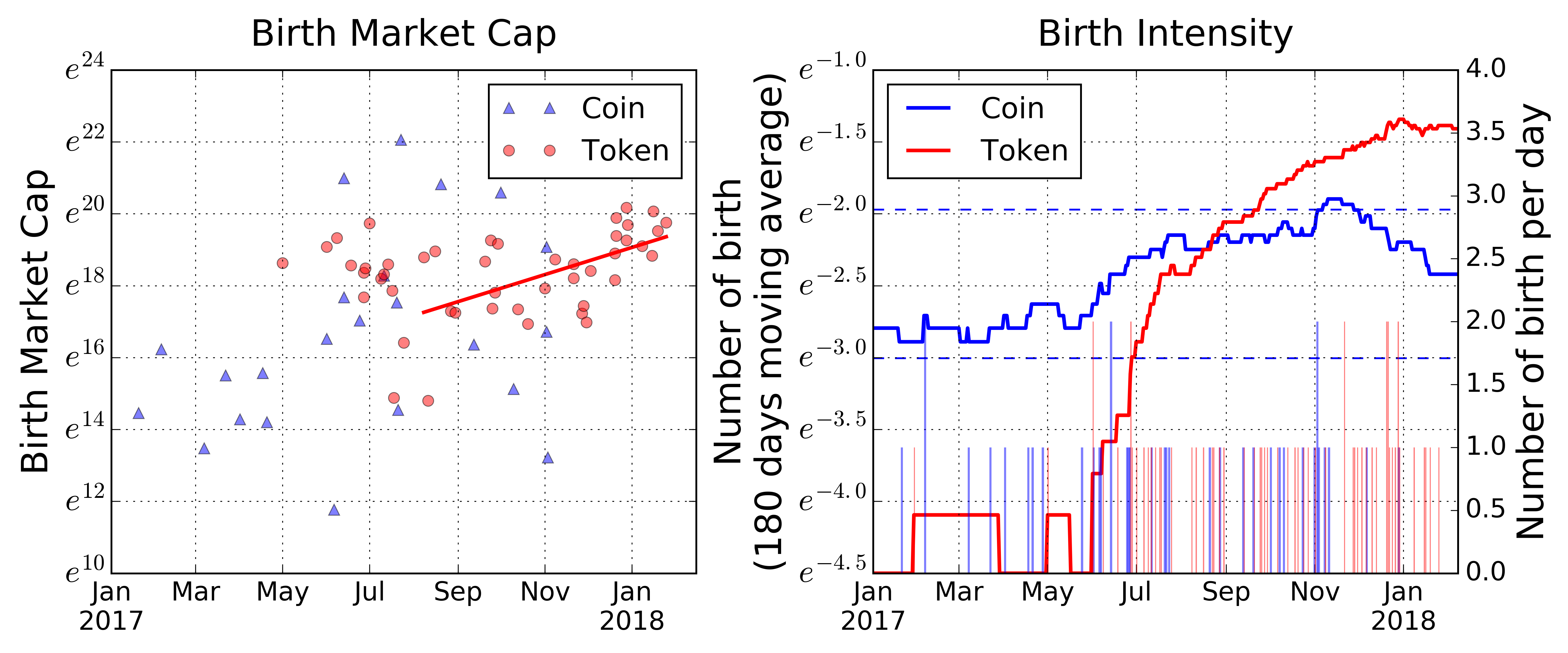}
	\caption{Birth market capitalization and birth intensity (frequency) of large coins (having average market capitalization over life larger than $10^{7.3}$) and tokens (average market capitalization over life time larger than $10^{8.1}$) since 2017. The left panel is the birth market capitalization of coins (blue triangles) and tokens (red circles). The red solid line is the best linear fit of the tokens' birth market capitalization. The right panel plots the number of birth of coins (blue) and tokens (red), smoothed by a 180 day moving average. The horizontal blue dashed lines in the middle and lower panels are the 95\% confidence intervals assuming a Poisson process for coins, whose mean is estimated on the full window (since 2017). The number of births per day is shown in blue (coins) and red (tokens) bars, against the right y-axis.}
	\label{fig:birth_size_intensity}
\end{figure}

Moving to the extended Gibrat's Law framework of Malevergne et al.~\cite{malevergne2013zipf}, the growth rate of birth size $c_0$, the growth of the birth intensity $d$, and the exit hazard rate $h$ are estimated, as summarized in Figure~\ref{fig:birth_size_intensity}. Importantly, as a threshold is necessary, only coins (resp. tokens) having an average market capitalization over their lifetime larger than US Dollars $10^{7.3}$ (resp. $10^{8.1}$) are considered, which correspond to roughly the top 10\% of coins and tokens. Further, due to non-stationarities not permitted by the framework, we focus on estimating the parameters in the most recent relatively stable window. In particular, the birth size of (high market capitalization) coins does not have a significant trend (p-value > 10\%), but rather shifted from one level to another around May 2017. Therefore, we fix the growth rate of birth size to be $c_0=0$. For tokens, however the birth size is significantly growing with time (p-value < 0.1\%), especially after July 2017. Thus, $c_0$ of tokens is estimated to be 1.19\%.  The birth intensity of coins is relatively stable, giving the growth rate of birth intensity $d=0$. However, the number of tokens has been growing significantly since May 2017 due to a large amount of ICOs (Initial Coin Offerings -- like an initial public offering of equity). A linear approximation of the growing birth intensity implies a Token birth intensity growth rate of $d$ to be 0.59\%.
In terms of the death process, there have been less than 3 dead large coins and tokens, so we consider the exit hazard rate $h$ to be 0 for both coins and tokens.\footnote{However, note that the empirical distributions of the lifetimes of all coins and tokens suggest similar Exponential distributions, indicating a similar death hazard rate for coins and tokens.}

\subsection{Comparing Empirical and Theoretical Predicted Distributions}

Given estimates of the five parameters in Equation~\ref{eqt:mu_th}, for coins and tokens separately, the theoretically predicted power law exponents are computed and compared with their empirical counterparts\footnote{Empirical exponents are based on the market capitalization distribution on Feb 7, 2018, as shown in Figure~\ref{fig:ccdf_initial_coin_token}.}, and summarized in Table~\ref{tab:prediction}.

Despite admitted model limitations, the empirically and theoretically predicted tail exponents of the market capitalization distributions are consistent, with the theoretical coin and token tail exponents, being less than and greater than 1 respectively. Comparing the empirical and theoretical tail exponents on time windows different from the one presented here is complicated by non-stationarities in the birth and death parameters, but still produces consistent results, with the theoretical and empirical tail exponents falling within similar ranges. This, in combination with the confirmation of Gibrat's law, effectively verifies the proposed model, delivering a robust insight into the underlying nature of the two fundamentally different coin and token ecosystems.

For coins, we have $r-h>d+c_0$ which means that the capitalisation growth (corrected for death) of existing ``entrenched incumbents'', such as Bitcoin, Ethereum, and so on, exceeds the growth of capitalisation of recent market entrants, of which there are relatively few. This inequality also theoretically implies a tail exponent $\mu$ less than 1 under the framework of \cite{malevergne2013zipf}. In contrast, the token market has the opposite feature, with  $r-h\leq d+c_0$, implying a thinner tail with $\mu\leq 1$. Indeed in the token market, the high rate of birth of tokens is the dominant feature driving the market, and the limited growth $r$ in excess of death $h$ restricts growth of older tokens in relative terms, leading to a market capitalization distribution that is slightly lighter tailed than Zipf's law and reflecting an immature system.

\begin{table}
	\centering
	\caption{\label{tab:prediction} Comparison between the theoretically predicted power law exponents $\mu_{TH}$ and the empirical exponents $\mu_{MLE}$, estimated by maximum likelihood, for coins and tokens respectively. The theoretical values $\mu_{TH}$ are given by Equation~\ref{eqt:mu_th} with estimated birth and death parameters plugged in -- see the previous subsections for their estimation. Numbers in brackets are the 95\% confidence interval estimated by bootstrap. }
	\begin{tabular}{|lll|}
		\hline
		& \textbf{Coin} & \textbf{Token} \\ \hline
		\textbf{Growth rate of market capitalization $r$}     & 1.52\% [1.45\%, 1.59\%]   & 0.77\%  [0.68\%, 0.86\%]\\ \hline
		\textbf{Growth volatility $\sigma$}             & 24.67\% [22.64\%, 26.58\%]  & 24.6\% [22.74\%, 26.45\%]         \\ \hline
		\textbf{Exit hazard rate $h$}              & 0             & 0              \\ \hline
		\textbf{Growth rate of birth size $c_0$}  & 0             & 1.19\% [0.48\%, 1.90\%] \\ \hline
		\textbf{Growth of the birth intensity $d$} & 0             & 0.59\% [0.57\%, 0.61\%] \\ \hline
		\textbf{Empirical tail exponent $\mu_{MLE}$}  & 0.48 [0.39, 0.57]   & 1.29 [0.83, 1.75]          \\ \hline
		\textbf{Theoretical tail exponent $\mu_{TH}$} & 0.50 [0.41, 0.57]    & 1.29 [1.12, 1.48] \\ \hline
	\end{tabular}
\end{table}

\section{Discussion}

Having looked at the market capitalization of all cryptocurrencies, and treating coins and tokens separately, we aimed to understand the basic growth mechanism with a simple model. We have empirically verified that, for large coins and tokens, their market capitalizations follow power law distributions with significantly different values -- with the tail exponent falling between 0.5 and 0.7 for coins, and between 1.0 and 1.3 for tokens. 

Despite recognized limitations, the simple stochastic proportional growth model of Malevergne et al. \cite{malevergne2013zipf} successfully recovers these tail exponents based on statistically estimated birth, death, and proportional growth parameters. This clearly characterizes coins as being ``entrenched incumbents'' and tokens as an ``explosive immature ecosystem'', largely due to massive and exuberant ICO activity in the token space. 

With Zipf's Law having a unit tail exponent and being a statistical signature of an optimal economy \cite{malevergne2013zipf}, it is perhaps unsurprising that the coin and token markets have different tail exponents. Undoubtedly, if more productive regulation is introduced \cite{finma2018ico,chuan2017china,higgins2017surprise,kim2017southkorea}, and institutional investors flood the market and adoption grows, the market will become more mature. One can then expect a better balance between the growth of incumbents and a healthy rate of new entrants associated with technological innovations. However, as the cryptocurrencies are evolving towards being an alternative investment asset, one should remain extremely cautious, where massive endogenous instabilities exist \cite{sornette2003stock} and risks are poorly understood.

Looking forward, the methodology presented here could be productively extended to allow for varying quality (i.e., fitness) of cryptocurrencies. This would be realistic, as improved technology enters the market in new coins. In particular, such a framework could more adequately address the question of if and when Bitcoin will be overtaken, as pure proportional growth frameworks perhaps overly emphasize the strength of the so-called ``first mover advantage''.

\bibliographystyle{unsrt}
\bibliography{crypto_mktcap}

\end{document}